\documentclass[runningheads]{llncs}
\usepackage{graphicx}
\usepackage{amsfonts}
\usepackage{color}
\usepackage{amsmath}
\usepackage{graphicx}
\usepackage{amssymb}
\usepackage{amsmath}
\usepackage{amsfonts}
\usepackage{color}
\usepackage{amsmath}
\usepackage[linesnumbered,ruled]{algorithm2e}

\def\L\mathcal{L}

\def\D{\mathbf{D}}

\def\x{\mathbf{x}}
\def\z{\mathbf{z}}

\def\G{\mathbf{G}}

\def\L{\mathcal{L}}
\def\E{\mathbf{E}}

\def\x{\mathbf{x}}

\def\z{\mathbf{z}}

\def\0{\mathbf{0}}

\def\G{{G}}
\def\E{E}

\def\L{\mathcal{L}}
\def\D{D}
\def\x{x}
\def\z{z}
\usepackage{lineno,hyperref}
\modulolinenumbers[5]

\usepackage{mwe} % to get dummy images
\usepackage{multirow}
\usepackage{color}
\usepackage{tabu, tabularx}

\begin{document}
\title{Cross-Domain Medical Image Translation by Shared Latent Gaussian Mixture Model}
\titlerunning{Cross-Domain Medical Image Translation}
%\titlerunning{Medical Image Translation by Shared Latent Gaussian Mixture Model}
% If the paper title is too long for the running head, you can set
% an abbreviated paper title here
%
%\author{Submission 1842}
%
\author{Yingying Zhu\inst{1}\and Youbao Tang\inst{1} \and Yuxing Tang\inst{1} \and Daniel C. Elton\inst{1} \and \\  Sungwon Lee\inst{1} \and Perry J. Pickhardt\inst{2} \and Ronald M. Summers\inst{1}}
\authorrunning{Zhu et al.}
% First names are abbreviated in the running head.
% If there are more than two authors, 'et al.' is used.
%
\institute{Imaging Biomarkers and Computer-Aided Diagnosis Laboratory, Radiology and Imaging Sciences, National Institutes of Health, Clinical Center, Bethesda, MD 20892, USA \\ \email{yingying.zhu@nih.gov, rms@nih.gov }
\and School of Medicine and Public Health, University of Wisconsin, Madison, WI 53706, USA } 
 %\and Springer Heidelberg, Tiergartenstr. 17, 69121 Heidelberg, Germany
% \email{lncs@springer.com}\\
% \url{http://www.springer.com/gp/computer-science/lncs} \and
% ABC Institute, Rupert-Karls-University Heidelberg, Heidelberg, Germany\\
% \email{yingying.zhu@uta.edu}}
%
\maketitle % typeset the header of the contribution
\begin{abstract}
Current deep learning based segmentation models often generalize poorly between domains due to insufficient training data. In real-world clinical applications, cross-domain image analysis tools are in high demand since medical images from different domains are often needed to achieve a precise diagnosis. An important example in radiology is generalizing from non-contrast CT to contrast enhanced CTs. Contrast enhanced CT scans at different phases are used to enhance certain pathologies or organs. 
% Many current image segmentation works proposed to use cross-domain image translation to achieve segmentation. Such works focus on segmenting large organs since current cross-domain image translation models show impressive results on preserving large structures across domains. However, these works lack the ability to preserve fine structures in images.
Many existing cross-domain image-to-image translation models have been shown to improve cross-domain segmentation of large organs. However, such models lack the ability to preserve fine structures during the translation process, which is significant for many clinical applications, such as segmenting small calcified plaques in the aorta and pelvic arteries. In order to preserve fine structures during medical image translation, we propose a patch-based model using shared latent variables from a Gaussian mixture model. We compare our image translation framework to several state-of-the-art methods on cross-domain image translation and show our model does a better job preserving fine structures. The superior performance of our model is verified by performing two tasks with the translated images - detection and segmentation of aortic plaques and pancreas segmentation. We expect the utility of our framework will extend to other problems beyond segmentation due to the improved quality of the generated images and enhanced ability to preserve small structures. 

% \keywords{image-to-image Translation \and Non-Contrast CT \and Contrast Enhanced CT \and Gaussian Mixture Model \and Calcified Plaque Segmentation \and Pancreas Segmentation}
\end{abstract}
\section{Introduction}

Developing deep learning based segmentation models which can generalize to different domains has been in high demand since different types of medical images are usually collected in real clinical practice to achieve a precise  diagnosis. For example, a patient might have a non-contrast and a contrast-enhanced CT scan generated by injecting an intravenous contrast agent to highlight different internal structures at different time points. For instance, the arteries are  enhanced in early the early phase and the kidneys are enhanced in the late phase as the contrast agent is metabolized in the kidneys.

Although many existing works \cite{tang2018ct,jin2018ct,Tang2019SPIE,tang2019abnormal,tang2019tuna,tang2019xlsor} have used image-to-image translation techniques to assist in medical image analysis tasks, less work been done to address cross-domain image segmentation due to the lack of sufficient labelled data from different domains. %Few studies has been proposed for cross-domain image segmentation using limited labelled data on CT and MRI scans by cross-domain image translation~\cite{qidou2020,CAI2019174}. 
Only a few studies have used synthetic images generated by cross-modality image-to-image translation (e. g., CT and MRI) for cross-domain image (e. g., organ) segmentation~\cite{taozhou2020,zhu2020image}. However, image segmentation across CT and MRI scans is of less clinical importance due to its poor performance caused by the large difference between the two modalities. Moreover, CT scans are typically used to scan a large range (fast but low resolution) while MRI scans are often targeted at small regions (slow but high resolution). Paired CT and MRI scans for the same region of the body are rarely collected in clinical practice. Such works usually focus on large organ (e. g., heart) segmentation. However, hardly any work has been done for the small structure segmentation across domains, for example  calcified plaque segmentation in the aorta and pelvic arteries under different contrast levels. This is a clinically important problem, since calcified plaque in the arteries is a strong predictor of heart attack~\cite{coronary19}.

\begin{figure}[t!]
 % Caption and label go in the first argument and the figure contents
 % go in the second argument
\label{fig:intro}
\centering
{{\includegraphics[width=.95\linewidth]{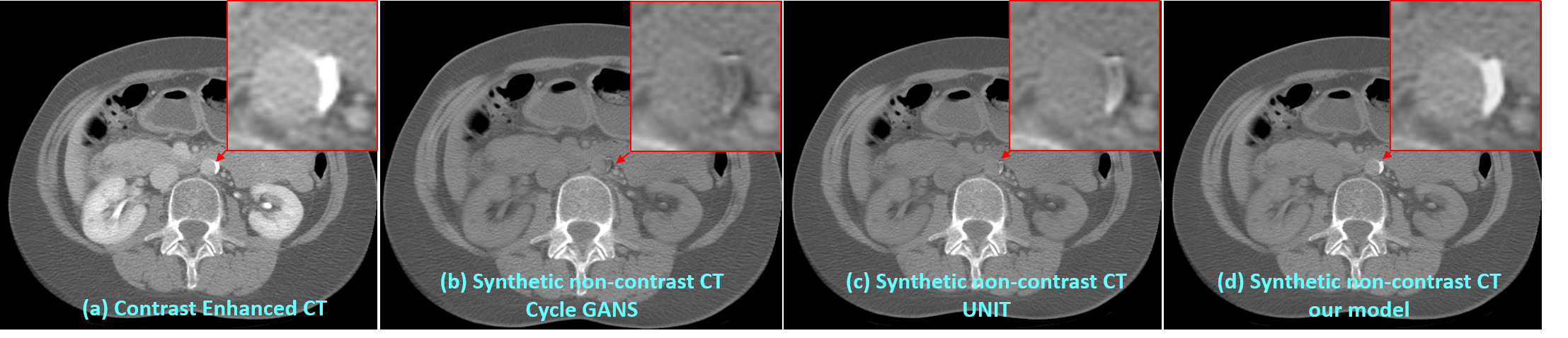}
  \caption{(a) Real post-contrast CT scan. (b) Synthetic non-contrast CT using CycleGAN~\cite{CycleGAN2017}.  (c) using shared latent variables from a Gaussian distribution~\cite{UNIT2017,huang2018munit}. (d) Gaussian mixture model.}}
  }

\end{figure}

Applying existing image translation models to improve calcified plaque segmentation on different domains is impractical since these model show inconsistent performance for preserving fine/tiny structures after image translation (as shown in Fig.~\ref{fig:intro}, the calcified plaques are blurry and covered by neighboring structures by CycleGAN \cite{CycleGAN2017} and UNIT~\cite{UNIT2017,huang2018munit}). We hypothesize this is because UNIT assumes a shared latent Gaussian variable across domains and real medical images actually lie in a shared Gaussian mixture model since different internal structures (image patches) lies in different local clusters as shown on the right side of Fig.~\ref{fig:intro}.

In order to address these problems, we proposed a patch-based domain invariant method using shared latent variables from a Gaussian mixture model for image translation~\cite{liu2018multi,Zhu-2014-17155}. In order to quantitatively evaluate the image translation performance of our model, we applied it to calcified plaque detection/segmentation and pancreas segmentation on both non-contrast and contrast enhanced CT images. We compared our model to several image translation networks. Experimental results showed that our model performs much better than the baseline (without image translation) and competing methods (e. g., CycleGAN~\cite{CycleGAN2017} and UNIT~\cite{UNIT2017}) by improving the performance of both tasks. {It is worth noting that our model is trained using unpaired images across different domains, which means it can be easily adapted to real-world clinical practice where paired images are relatively rare.}

%A common idea used in many image translation networks is that images from different domains share a latent single subspace and one can transfer between domains using this domain invariant latent space. Our method uses patches of size 32x32, each of which will lie in one or more subspace. Intuitively these subspaces may be particular organs, bones, and other body parts which are in the patch or of which the patch is a part. Recent work has demonstrated that enforcing that the latent space take the form of a union of subspaces can improve image reconstruction/generation on natural images especially when it comes to preserving fine structures~\cite{yujunshen2019,subgan18,junjianzhang19,Zhou_2018_CVPR}. We propose to modify the ``UNIT'' image translation network of \cite{UNIT2017} using a self-expressiveness loss, which results in subspace clustering in the latent space~\cite{NIPSSCN}. We show that our model can preserve subtle structures (such as calcified plaque) in cross-domain image translation compared to state-of-art methods~(shown in Fig.~\ref{fig:intro} right). We further applied this model to assist in a cross-domain calcified plaque segmentation task using a Mask-RCNN based model by~\cite{LiuSPIE2019}. Our model shows significant performance improvement compared to the state-of-the-art UNIT method~\cite{UNIT2017}. 
%It is worth noting that our approach can be generalized to different image translation networks and other types of images besides CT scans.

%----------------------------------------------------------- 
\section{Method}

\subsection{Unsupervised Image-to-Image Translation Networks (UNIT) }
{Let $\x_1 \in \mathcal{X}_1$ and $\x_2\in \mathcal{X}_2$ be two images from non-contrast and contrast enhanced CT, respectively. The image size of $\x_1,\x_2$ is $512\times 512$ in our dataset. Liu et al.~\cite{UNIT2017} proposed two sets of variational auto-encoders for the two different domains and translate the images across the two domains via a shared latent space $\mathcal{Z}$.}
{The shared latent space $\mathcal{Z}$ is conditionally independent from the two domains and is enforced to follow a Gaussian distribution with unit variance. Intuitively, this latent space $\mathcal{Z}$ encodes the underlying morphological structure of objects and is domain invariant. For example, the latent space $\mathcal{Z}$ may depend on the shape of internal organs which is invariant across image domains. They implemented this by sharing the latent space $\mathcal{Z}$ layers of two variational autoencoders. The UNIT model~\cite{UNIT2017} trained using the whole image, which results in loss of detailed structures as shown in Fig. 1(b). Similarly, the CycleGAN method also translates imags between domains using a shared latent space and cycle consistent loss but can not preserve the detailed structures (see Fig. 1(c)). }

\subsection{Patch-Based Mixtures Gaussian Image-to-Image Translation}
\begin{figure}[t!]
\label{fig:intro1}
\centering
{{\includegraphics[width=0.75\linewidth]{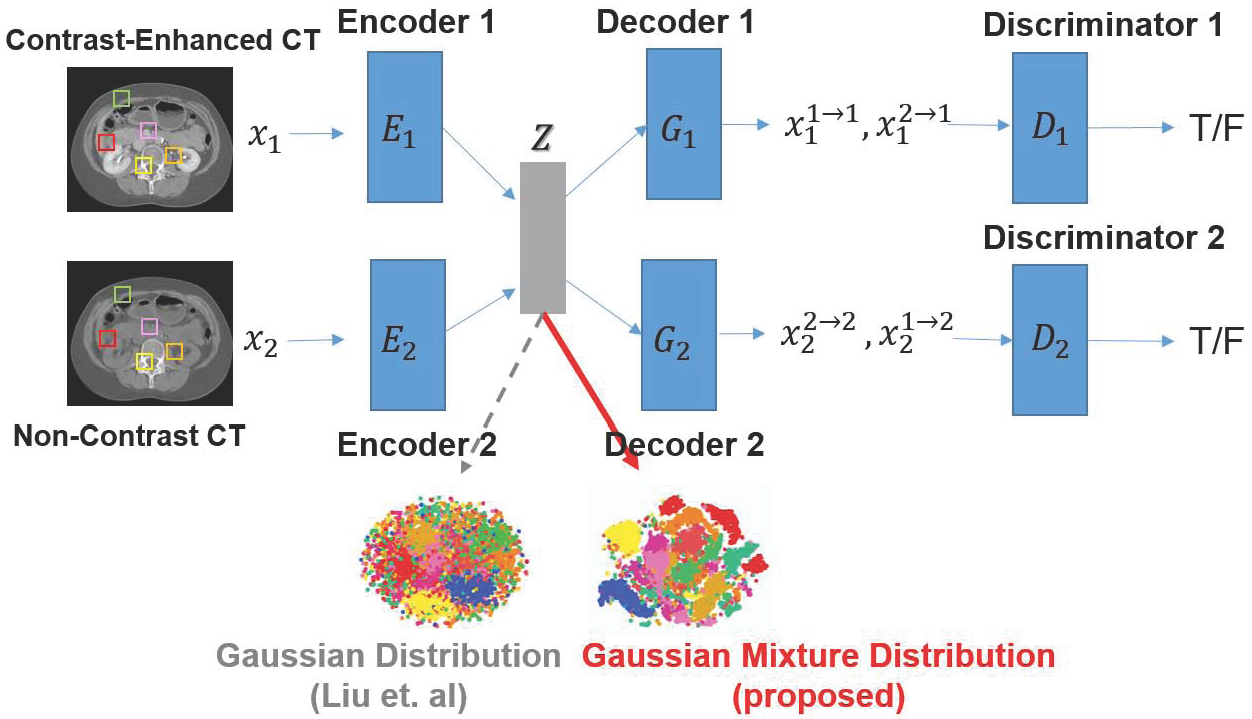}
\caption{The framework of our proposed model compared to UNIT~\cite{UNIT2017} model. Our model assumes that the shared latent variable lies in a Gaussian mixed distribution (different patches lies in different local clusters and Liu et. al. \cite{UNIT2017} assume a single Gaussian distribution.}
  }}
\end{figure}

\textcolor{black}{We proposed a patch based method and extracted many small randomly sampled patches from each image. The optimal patch size is determined using testing with the validation set. We used a patch size of $32\times 32$ in all our experiments. We extracted random patches at the same location from $\x_1,\x_2$ respectively. We extracted the image features using a pre-trained CNN on these small patches and found that they lie in different local clusters as shown in Fig. 2. Intuitively, image patches from different organs or internal structures will be clustered into different local clusters. Based on these observations, we proposed to model the domain independent shared latent space $\mathcal{Z}$ using a mixture Gaussian model, $\mathbf{\mu}_z = [\mu_{z,1},\cdots,\mu_{z,k}], \mathbf{\Sigma}_z=\mathbf{[\Sigma_{1,z},\cdots,\Sigma_{z,K}]}$ are the mean and variance for different Gaussian components.}
\begin{eqnarray}
\z\sim \sum_{k=1}^K \pi_k \mathcal{N}(z|\mu_{k,z}, \mathbf{\Sigma}_{k,z} ), s. t. \sum_{k=1}^K \pi_k = 1
\end{eqnarray},
\textcolor{black}{$K$ is the number of Gaussian components following ~\cite{Dilokthanakul2017DeepUC}. It is worth noting that $K$ is determined by the validation dataset on the downstream segmentation/detection task. We follow \cite{UNIT2017} and use 6 sub-networks: two domain image encoders: $\E_1, \E_2$, two domain image generators $\G_1,\G_2$, and two domain adversarial discriminators $\D_1,\D_2$ as shown in Fig.\ 2. We use VGG-16 as the encoder and a reversed VGG-16 structure as the decoder. We use the variational inference model from \textit{Edward} \cite{tran2016edward} to solve the parameters in our model. The encoder outputs a set of mean vectors for each Gaussian component: $E_1(\x_1,\theta_1), E_2(\x_2,\theta_2)$, $\mathbf{\Sigma_1 =[\Sigma_{1,1},\cdots,\Sigma_{1,K}}], 
\mathbf{\Sigma_{2} =[\Sigma_{2,1},\cdots,\Sigma_{2,K}}] $ are the variance matrix for each Gaussian component. $\mathbf{\Theta_1=[\theta_{1,1},\cdots,\theta_{1,K}]}$,
$\mathbf{\Theta_2 =[\theta_{2,1},\dots,\theta_{2,K}}$ are the weights for using a shared encoder to output the mean for different Gaussian components. 
The distribution of latent code $z$ given $x_1, x_2$ are listed as,}
\begin{eqnarray}
   \nonumber q_1({\z}|\x_1)  \sim  \sum_{i=1}^{K} \pi_k\mathcal{N}\left(
    \z|{E}_1(\x_1,\mathbf{\theta}_{1,k}) , \mathbf{\Sigma}_{k,1}) \right), \sum_{k=1}^K \pi_k= 1,\\
 \nonumber q_2({\z}|\x_2)  \sim  \sum_{i=1}^{K} \pi_k \mathcal{N}\left(
    \z|{E}_2(\x_2,
    \mathbf{\theta}_{2,k}) ,\mathbf{\Sigma}_{k,2}) \right), \sum_{k=1}^K \pi_k= 1.
\end{eqnarray}

%\begin{equation}
%    p({\x}_2|\z) = \sum_{i=1}^{K} \pi_i(\x_2)\mathcal{N}(\mathbf{\mu}_i(\x_2) ,\mathbf{\sigma}_i(\x_2) ), \ s. t. \sum_{i}^K{\pi_i({\x_2})} = 1
%\end{equation}
\color{black}{The reconstructed image $\hat{x}_1^{1->1} =G_1(z\sim q_1(z|x_1))$ for two variational autoenconder (VAE) $(E_1,G_1)$ and $(E_2,G_2)$ are $\hat{x}_2^{2->2} =G_2(z\sim q_2(z|x_2))$, we have the VAE loss:}
\begin{eqnarray}
&&\mathcal{L}_{VAE_1}(E_1,G_1,\mathbf{\Theta}_1,\mathbf{\Sigma}_1,\mathbf{\Sigma}_z,\mathbf{\mu}_{z}) = \lambda_1 \mbox{KL}(q_1(z|x_1)||p(z)) \\
\nonumber&& -\lambda_2 \mathbb{E}_{z\sim q_1(z|x_1)}[\log pG_1(x_1|z)]\\
&&\mathcal{L}_{VAE_2}(E_2,G_2,\mathbf{\Theta}_2,\mathbf{\Sigma}_2,\mathbf{\Sigma}_z,\mathbf{\mu}_{z}) = \lambda_1 \mbox{KL}(q_2(z|x_2)||p(z))\\
\nonumber &&- \lambda_2 \mathbb{E}_{z\sim q_2(z|x_2)}[\log pG_1(x_2|z)],
\end{eqnarray}
\textcolor{black}{where $\lambda_1,\lambda_2$ are the parameters controls the weights for the objective terms and the KL divergence terms penalized derivation of the distribution if the latent variable from the prior distribution.}

\textcolor{black}{Our model also has two generative adversarial networks: $GAN_1 = \{G_1, D_1\}$, $GAN_2 = \{G_2,D_2\}$. $D_1, D_2$ are constrained to output true if images are sampled from the first or second domain respectively and output false if the images are generated from $G_1, G_2$ respectively. We have the following conditional GAN objective functions, which constrain the translated images to resemble images in their respective target domains:}
\begin{eqnarray}
&&\mathcal{L}_{GAN_1}(E_2,G_1,D_1,\mathbf{\Theta}_1,\mathbf{\Sigma}_1,\mathbf{\Sigma}_z,\mathbf{\mu}_{z}) = \lambda_0 \mathbb{E}_{x_1\sim P_{\mathcal{X}_1}} [\log 
D_1(x_1) ]\\
\nonumber &&+ \lambda_0\mathbb{E}_{z\sim q_2(z|x_2)}[\log D_1(G_1(z))]\\
&&\mathcal{L}_{GAN_2}(E_1,G_2,D_2,\mathbf{\Theta}_2,\mathbf{\Sigma}_2,\mathbf{\Sigma}_z,\mathbf{\mu}_{z}) = \lambda_0 \mathbb{E}_{x_2\sim P_{\mathcal{X}_2}} [\log D_2(x_2) ] \\
\nonumber &&+ \lambda_0\mathbb{E}_{z\sim q_1(z|x_1)}[\log D_2(G_2(z))],
\end{eqnarray}
\textcolor{black}{Similarily to the previous equation, the hyperparameter $\lambda_0$ balances the impact of the $GAN$ objective function. We also incorporate a cycle consistency constraint to ensure that twice translated images resemble the original image and a KL divergence term which penalizes the latent code from deviating too far from the prior distribution.}
\begin{eqnarray}
&&\mathcal{L}_{CC_2}(E_1,G_1,E_2,G_2,\mathbf{\Theta}_1,\mathbf{\Theta}_2, \mathbf{\Sigma}_1, \mathbf{\Sigma}_2,\mathbf{\Sigma}_z,\mathbf{\mu}_{z} )\\
&& \nonumber = \lambda_3 \mbox{KL}(q_1(z|x_1)||p(z))\\
\nonumber &&+ \lambda_4\mbox{KL}(q_2(z|x_1^{1->2})||p(z)) \nonumber- \lambda_4 \mathbb{E}_{z\sim q_2(z|x_1^{1->2})}[\log pG_1(x_1|z)]\\
\nonumber&&\mathcal{L}_{CC_2}(E_2,G_2,E_1,G_1,\mathbf{\Theta}_1,\mathbf{\Theta}_2,\mathbf{\Sigma}_1,\mathbf{\Sigma}_2,\mathbf{\Sigma}_z,\mathbf{\mu}_{z}) \\
&&= \lambda_3 \mbox{KL}(q_2(z|x_2)||p(z)) \\
\nonumber &&+ \lambda_4 \mbox{KL}(q_1(z|x_2^{2->1})||p(z)) 
- \lambda_4 \mathbb{E}_{z\sim q_1(z|x_2^{2->1})}[\log pG_2(x_2|z)]
\end{eqnarray}

\textcolor{black}{Combining all the above objective functions, our final objective functions is:}

\begin{eqnarray}
&&\arg\min(\E_1,\E_2,\G_1,\G_2,\mathbf{\Theta}_1, \mathbf{\Theta}_2,\mathbf{\Sigma}_1,\mathbf{\Sigma}_2,\mathbf{\Sigma}_z,\mathbf{\mu}_{z}) \max(\D_1,\D_2) \\
\nonumber&&\mathcal{L}_{VAE_1}(\E_1,\G_1,\mathbf{\Theta_1,\Sigma_1},\mathbf{\Sigma}_z,\mathbf{\mu}_{z}) +\mathcal{L}_{VAE_2}(\E_2,\G_2,\mathbf{\Theta_1,\Sigma_1},\mathbf{\Sigma}_z,\mathbf{\mu}_{z})\\
&&+\nonumber\mathcal{L}_{CC1}(\E_1,\G_1,\E_2,\G_2,\mathbf{\Theta_1,\Sigma_1, \Theta_2,\Sigma_2},\mathbf{\Sigma}_z,\mathbf{\mu}_{z})\\
\nonumber &&+ \mathcal{L}_{CC2}(\E_1,\G_1,\E_2,\G_2,\mathbf{\Theta_1,\Sigma_1,\Theta_2,\Sigma_2},\mathbf{\Sigma}_z,\mathbf{\mu}_{z}) \\
\nonumber&&+ \mathcal{L}_{GAN_1}(\E_1,\G_1,\D_1,\mathbf{\Theta_1,\Sigma_1}) +\mathcal{L}_{GAN_2}(\E_2,\G_2,\D_2,\mathbf{\Theta_2,\Sigma_2})
\end{eqnarray}

%\textcolor{red}{We used Adam as our optimizer with an initial learning rate of 0.00001, and the learning rate decayed by half at every 50 epochs.}
%We fixed the trade-off coefficient λ1 in Eq.12, λ2 λ3 and λ4 to be 10,1,1 and 0.5. %Specifically, we firstly the image as 512, then crop the patch with the size as 256 from the whole image randomly.

%------------------------------------------------------ 
\section{Experiments}
%\textbf{Implementation.}
%The entire pipeline was implemented by PyTorch on a work-
%station equipped with NVIDIA RTX 2080 Ti. We used Adam as our optimizer with an initial learning rate of 0.00001, and the learning rate decayed by half at every 50 epochs. We used the warm-up weight to avoid the gradient exploding problem. 
%We fixed the trade-off coefficient λ1 in Eq.12, λ2 λ3 and λ4 to be 10,1,1 and 0.5. %Specifically, we firstly the image as 512, then crop the patch with the size as 256 from the whole image randomly.

\textbf{Evaluation tasks.} 
In order to evaluate the image translation model quantitatively, we evaluated the performance of our image translation models on two challenging image segmentation tasks: calcified plaque detection/segmentation and pancreas segmentation on both the non-contrast and contrast enhanced CT scans given labelled CT scans only from one domain. We also compared our model to two recent state-of-the art image translation models: CycleGAN~\cite{CycleGAN2017} and UNIT~\cite{UNIT2017,huang2018munit}, which have been broadly applied to image translation on natural images.

\textbf{Image Translation Training.} 
For training we used 140 unpaired CT scans (70 non-contrast, 70 contrast enhanced) taken from renal donor patients at the University of Wisconsin Medical Center. We applied our image translation model to generate synthetic contrast enhanced CT scans from the labelled non-contrast CT scans and used them as augmented data to improve our plaque segmentation/detection performance and pancreas segmentation on both contrast enhanced and non-contrast CT scans.  

\textcolor{black}{The image translation training dataset was separated into 10 folds and one fold was used as validation data for selecting hyperparameters including the number of Gaussian mixture components $K$ and $\lambda_0,\lambda_1,\lambda_2,\lambda_3,\lambda_4$. The validation dataset was evaluated for downstream plaque detection tasks using the model trained from section 3.1. The optimal value of $K$ is very important for our final image translation results. A small $K$ value can lead to blurring of local image structures and a large $K$ value can add to  computation cost and more uncertainty on the output images. We used $K=25$ based on the validation dataset performance. It is worth noting that the setting of $K$ can vary between datasets. In order to select the best generated image for cross-domain image segmentation task, we also train a quality control network using an independent CT dataset selected from DeepLesion data~\cite{deeplesion} to remove the synthetic CT images with artifacts.}

\subsection{Calcified plaque detection and segmentation}

Labelling calcified plaques in the CT scans is very time consuming since the plaques are very small, frequently on the order of just a few pixels in CT images. In our experiments, we only have labelled training CT images from low dose non-contrast CT scans. We trained a 2D detection and segmentation model~\cite{Liu2019ISBI} on 75 low dose CT scans which contained a total of 25,200 images (transverse cross sections/slices), including 2,119 with plaques. The training dataset was divided into 10 folds and we used 9 folds as the training dataset and 1 fold as the validation dataset for parameter selection. We shuffled the training and validation dataset and trained 10 2D Mask R-CNN~\cite{maskrcnn} models and applied these models to our independent testing dataset. We report the mean and standard derivation across all 10 models in Table 1. For this work we labelled an additional testing dataset with 30 contrast enhanced CT scans and 30 non-contrast scans from a different dataset collected at University of Wisconsin Medical Center. It had plaque labeled manually (7/30 of these scans contained aortic plaques, with a total of 53 plaques overall). 

\begin{figure}[t!]
\label{fig:intro1}
\centering
\includegraphics[width=0.9\linewidth]{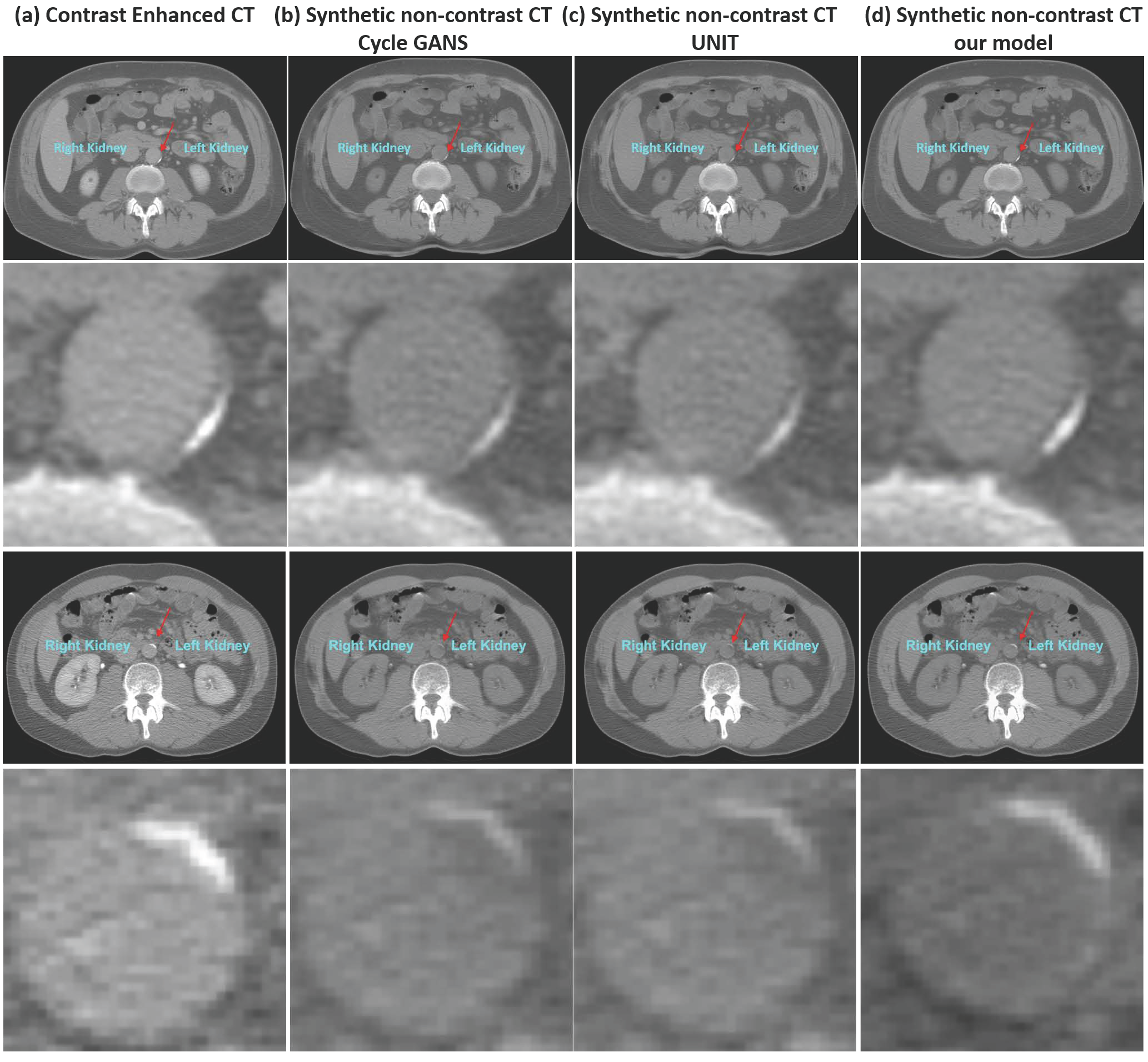}
\caption{Visualization of synthetic non-contrast CT scans from contrast enhanced CT (a) scans by (b) CycleGAN (c) UNIT (d) our model. We show two cases of median quality generated images by all competing methods. }
\end{figure}

We selected an synthetic image which shows the median performance on image translation (as shown in Fig. 3). Fig. 3 (a) shows contrast enhanced CT scans in the late phase (the kidneys are enhanced with bright pixel values). There is a very small plaque in this image which was translated into non-contrast CT scans by cycleGAN~\cite{CycleGAN2017}, UNIT~\cite{UNIT2017} and our model. The whole translated images looks very similar by all competing methods, however, the calcified plaque pixel brightness is better preserved by our approach.  
%{As shown in Fig.~\ref{fig:intro}, (a) is a real post-contrast CT scan in the late phase (contrast are showing majorly in kidneys). (b) show that Liu et al.'s method remove the contrast at the cost of losing fine plaque structures. (c) show that our model removes the contrast in the kidneys and preserves the fine calcified plaques. } 

\begin{table}[t!]
\centering
\caption{Plaque detection and segmentation results. The first column gives detection results for the original data without image translation. The 2nd, 3rd, and 4th columns give results for non-contrast (NC) and contrast enhanced (CE)  plaque detection after non-contrast to contrast enhanced image translation with different image translation models. SYN= synthetic images.}
\scriptsize
\begin{tabu} to 0.995\textwidth {| X[0.8,c] | X[1,c] | X[1.2,c] | X[1.2,c] | X[1.3,c] |}
\hline
Method & Baseline &  CycleGAN~\cite{Clark2013CIA} & UNIT~\cite{UNIT2017,huang2018munit} & Ours\\ 
 \hline
 \hline
Training  & NC   &  NC \& {SYN CE } &  { NC \& SYN CE }&{  NC \& SYN CE } \\

\hline
\hline
Testing &NC & NC  & NC  & NC  \\ 
 \hline
 Precision & 78.4 $\pm$ 1.82\%  & 79.5 $\pm$ 1.75\%  &  80.2 $\pm $ 1.68\%  & \textbf{80.7 $\pm $ 1.65\% }\\ 
 Recall &  82.4 $\pm $ 2.45\%   & 83.1 $\pm $ 2.26\%   &84.7 $\pm$ 2.13\%  & \textbf{85.2 $\pm$ 2.03}\%  \\ 
 Dice &  0.724 $\pm$0.245  & 0.733$\pm$0.223  &0.751 $\pm$ 0.218 & \textbf{0.756 $\pm$0.193 } \\ 
 \hline
 \hline
Testing &CE & CE  & CE  & CE  \\ 
\hline
 Precision &  48.6$\pm$ 3.52 \% &  61.5$\pm$ 2.87\% &   64.7 $\pm$ 2.64\%  &  \textbf{78.2}$\pm$ 2.58\%\\ 
 Recall &   54.3 $\pm $ 3.64\%  &  64.3$\pm $3.21\%  &69.8 $\pm$ 3.05\%  &  \textbf{81.2 $\pm$ 2.87}\% \\ 
 Dice &  0.452$\pm$0.251  &  0.534 $\pm$ 0.236 & 0.566 $\pm$0.198 & \textbf{0.676\ $\pm$0.176} \\ 
 \hline
\end{tabu}

\end{table}

\textbf{Quantitative Results.}
The calcified plaque detection and segmentation results are shown in Table 1. Our model achieved similar plaque detection and segmentation performance to the real pre-contrast CT scans (precision decreased about $>1.5\%$, recall decreased about $>4\%$) and dice coefficients drops about $>0.1$. The detection and segmentation model trained on synthetic images generated from UNIT~\cite{UNIT2017} and CycleGAN~\cite{CycleGAN2017}, by contrast, shows a $>15\%$ drop in precision, a $>13\%$ drop in recall and drop $>0.18$ in Dice coefficients caused by loss of fine structures. 
%Developing domain adaption model using minimum training data for medical imaging is critical for real world clinical application scenario since collecting large size medical dataset is very time consuming and expensive.

%Interestingly, we obtain similar improvement even when the number of training examples is cut from 140 to 60, showing that this method can be applied even when the number of training data is low. Future work may explore the application of this approach to other domains (such as different contrast phases) and cross domain disease diagnosis.  

%------------------------------------------------------ 
%\subsection{Calcified Plaque Detection and Segmentation}
%We use the developed Maskrcnn tools from~\cite{LiuSPIE2019} for calcified plaque detection and segmentation. We trained it on a data set from ** with *** pre-contrast CT scans labelled with calcified plaques and applied it on the synthetic pre-contrast CT scans from the testing dataset 

%\textcolor{red}{Plaque segmentation with/without image translation (UNIT/our model) I am not sure whether we need this if it is not that great.  The overall idea is to show that we modified the image translation model to get better images. }

%------------------------------------------------------ 
%\subsubsection{UNet Segmentation Results \textcolor{red}{If it works, we add this section, Daniel will do this section}}

%---------------------------------------------------------------------------------------

\subsection{Pancreas segmentation}

Pancreas segmentation is very important for the diagnosis of pancreas cancer and surgical planning. Pancreas segmentation is challenging since the pancreas is very small compared to other internal organs and has large variance in its shape and orientation. Most existing pancreas segmentation approaches focus on segmenting pancreas only in contrast enhanced CT where the pancreas structures are more enhanced and have clearer boundaries. Current public pancreas segmentation data are only labelled on contrast enhanced CT.
We combined two public contrast enhanced CT datasets for pancreas segmentation. The first one includes 82 labelled contrast enhanced CT scans from the Cancer Imaging Archive database and second one has 281 contrast enhanced CT scans from the Medical Segmentation Decathlon database~\cite{pancreas2,pancreas15}. 

We use 10-fold cross validation and report the mean and standard derivation across the 10 folds. In order to improve pancreas segmentation on non-contrast CT images, we generated non-contrast CT from these contrast enhanced CT and used them to train a cross-domain 3D segmentation model. 
We use the multiple scale 3D Unet model proposed in~\cite{SaiSPIE2019} and compared with a 3D U-Net trained using synthetic non-contrast CT scans generated the different image translation models (CycleGAN~\cite{CycleGAN2017}, UNIT~\cite{UNIT2017,huang2018munit} and our model). 
We use 24 non-contrast CT scans annotated by an expert radiologist as the non-contrast pancreas segmentation testing dataset.

\begin{table}[t!]
\centering
\caption{Pancreas segmentation results with false positives pixel numbers and Dice scores on contrast enhanced CT and non-contrast CT. The first column show the baseline model which is only trained using contrast enhanced CT scans. Th second, third, and last column show the results trained by contrast enhanced CT and synthetic non-contrast CT generated by CycleGAN~\cite{CycleGAN2017}, UNIT~\cite{UNIT2017,huang2018munit} and our method respectively. NC=non-contrast CT. CE=contrast enhanced CT. SYN=synthetic. }
\scriptsize
\begin{tabu}to 0.995\textwidth {| X[0.6,c] | X[0.9,c] | X[1.3,c] | X[1.4,c] | X[1.2,c] |}
\hline
Method & Baseline &  CycleGAN \cite{CycleGAN2017} & UNIT \cite{UNIT2017,huang2018munit} & Ours\\
 \hline  \hline
Training  & CE   &  CE \& {SYN NC} &  { CE \& SYN NC  }&{  CE \& SYN NC } \\
\hline
\hline
Testing &CE & CE  & CE  & CE  \\ 
 \hline
%  FP &  37856 $\pm$ 30152    & 37235$\pm$ 30324   & 36824$\pm$ 31052  & \textbf{ 36351$\pm$ 30912} \\ 
Precision &   85.7$\pm$ 1.82 \%  &  87.2$\pm$1.62\%  &  88.1$\pm$ 1.53\%  &  \textbf{89.5 $\pm$ 1.44 \%} \\ 
Recall &   86.8$\pm$ 2.1\%  &  88.7$\pm$1.97\%  &  89.5$\pm$ 1.83\%  &  \textbf{90.7 $\pm$ 1.68\%} \\ 
Dice &  0.728$\pm$ 0.173   & 0.728$\pm$ 0.154   &0.731 $\pm$ 0.142  & \textbf{0.734 $\pm$ 0.136}  \\ 
  \hline
\hline
Testing &NC & NC  & NC  & NC  \\ 
 
 \hline
 % FP &   84506$\pm$ 43238  &  42204$\pm$28421  &  38924$\pm$ 30176  &  \textbf{34526 $\pm$ 27908} \\ 
Precision &   78.1$\pm$ 2.83 \%  &  82.8$\pm$2.67\%  &  83.2$\pm$ 2.45\%  &  \textbf{84.7 $\pm$ 2.25 \%} \\ 
Recall &   81.5$\pm$ 3.0\%  &  84.3$\pm$2.81\%  &  86.2$\pm$ 2.75\%  &  \textbf{87.2 $\pm$ 2.56\%} \\ 
Dice &  0.642 $\pm $ 0.183  & 0.684 $\pm$ 0.172  & 0.697 $\pm$ 0.163 &  \textbf{0.725$\pm$0.153} \\ 
\hline
\end{tabu}

\label{TB:pancreas}
\end{table}

\textbf{Quantitative results.} 
Table ~\ref{TB:pancreas} shows the quantitative results of cross domain pancreas segmentation results on both contrast enhanced CT and non-contrast CT using different image translation methods (CycleGAN~\cite{CycleGAN2017}, UNIT~\cite{UNIT2017} and our model). As shown in the top part of the table, the pancreas segmentation results on contrast enhanced CT scans are similar for all competing methods. Adding synthetic non-contrast CT scans in the training can also improve the pancreas segmentation on contrast enhanced CT, and our method shows slight improvement compared to all other methods. For pancreas segmentation on non-contrast CT, adding synthetic non-contrast CT images in the training can significantly improve the segmentation Dice score and reduce false positive pixels. For example, CycleGAN and UNIT show an average improvement of Dice score $>0.04/>0.05$ and reduction in false positive pixels $>4000/>4500$ compared to the baseline model. Our model shows the best performance and achieves a $>0.08$ improvement in Dice score and a reduction of  $>5000$ false positive pixels on average compared to the baseline model. 

\section{Conclusion}
In this work, we proposed an image translation model using shared latent variables from a Gaussian mixture distribution to preserve fine structures on medical images. We applied our method to two challenging medical imaging segmentation tasks: cross domain (non-contrast and contrast enhanced CT) calcified plaque detection/segmentation and pancreas segmentation. We demonstrated that our model can translate the medical images across different domains with better preservation of fine structures compared to two state-of-the-art image translation models for natural images. 
In the future work, we will explore the application of this model to translating medical images across multiple domains, for example contrast enhanced CT scans at different phases from non-contrast CT scans. Possible applications of this method are generating synthetic images to reduce radiation dose or creating 100\% contrast enhanced CT scans from $\approx$10\% dose contrast enhanced CT scans to reduce the dose of intravenous contrast agent used on patients.

\section{Acknowledgments}
This research was supported in part by the Intramural Research Program of the National Institutes of Health Clinical Center. We thank NVIDIA for GPU card donations.

%----------------------------------------------------------------------- 
%
% ---- Bibliography ----
%
% BibTeX users should specify bibliography style 'splncs04'.
% References will then be sorted and formatted in the correct style.
%
\bibliographystyle{splncs04}
\bibliography{ref}

\begin{thebibliography}{10}
\providecommand{\url}[1]{\texttt{#1}}
\providecommand{\urlprefix}{URL }
\providecommand{\doi}[1]{https://doi.org/#1}

\bibitem{Clark2013CIA}
Clark, K., Vendt, B., Smith, K., Freymann, J., Kirby, J., Koppel, P., Moore,
  S., Phillips, S., Maffitt, D., Pringle, M., et~al.: The cancer imaging
  archive (tcia): maintaining and operating a public information repository.
  Journal of Digital Imaging  \textbf{26}(6),  1045--1057 (2013)

\bibitem{Dilokthanakul2017DeepUC}
Dilokthanakul, N., Mediano, P.A., Garnelo, M., Lee, M.C., Salimbeni, H.,
  Arulkumaran, K., Shanahan, M.: Deep unsupervised clustering with gaussian
  mixture variational autoencoders. arXiv e-prints:1611.02648  (2016)

\bibitem{maskrcnn}
{He}, K., {Gkioxari}, G., {Dollár}, P., {Girshick}, R.: Mask {R-CNN}. In:
  International Conference on Computer Vision (ICCV). pp. 2980--2988 (2017)

\bibitem{huang2018munit}
Huang, X., Liu, M.Y., Belongie, S., Kautz, J.: Multimodal unsupervised
  image-to-image translation. In: European Conference on Computer Vision
  (ECCV). pp. 172--189 (2018)

\bibitem{jin2018ct}
Jin, D., Xu, Z., Tang, Y., Harrison, A.P., Mollura, D.J.: Ct-realistic lung
  nodule simulation from 3{D} conditional generative adversarial networks for
  robust lung segmentation. In: International Conference on Medical Image
  Computing and Computer-Assisted Intervention. pp. 732--740 (2018)

\bibitem{Liu2019ISBI}
Liu, J., Yao, J., Bagheri, M., Sandfort, V., Summers, R.M.: A semi-supervised
  {CNN} learning method with pseudo-class labels for atherosclerotic vascular
  calcification detection. In: International Symposium on Biomedical Imaging
  (ISBI). pp. 780--783 (2019)

\bibitem{liu2018multi}
Liu, L., Nie, F., Wiliem, A., Li, Z., Zhang, T., Lovell, B.C.: Multi-modal
  joint clustering with application for unsupervised attribute discovery. IEEE
  Transactions on Image Processing  \textbf{27}(9),  4345--4356 (2018)

\bibitem{UNIT2017}
Liu, M.Y., Breuel, T., Kautz, J.: Unsupervised image-to-image translation
  networks. In: Advances in neural information processing systems (NeurIPS).
  pp. 700--708 (2017)

\bibitem{coronary19}
Parab, S.Y., Patil, V.P., Shetmahajan, M., Kanaparthi, A.: Coronary artery
  calcification on chest computed tomography scan--anaesthetic implications.
  Indian Journal of Anaesthesia  \textbf{63}(8), ~663 (2019)

\bibitem{pancreas15}
Roth, H.R., Lu, L., Farag, A., Shin, H.C., Liu, J., Turkbey, E.B., Summers,
  R.M.: Deeporgan: Multi-level deep convolutional networks for automated
  pancreas segmentation. In: International Conference on Medical Image
  Computing and Computer-Assisted Intervention (MICCAI). pp. 556--564 (2015)

\bibitem{pancreas2}
Simpson, A.L., Antonelli, M., Bakas, S., Bilello, M., Farahani, K.,
  Van~Ginneken, B., Kopp-Schneider, A., Landman, B.A., Litjens, G., Menze, B.,
  et~al.: A large annotated medical image dataset for the development and
  evaluation of segmentation algorithms. arXiv e-prints: 1902.09063  (2019)

\bibitem{SaiSPIE2019}
Sriram, S.A., Paul, A., Zhu, Y., Sandfort, V., Pickhardt, P.J., Summers, R.:
  Multilevel {U-Net} for pancreas segmentation from non-contrast {CT} scans
  through domain adaptation. In: Hahn, H.K., Mazurowski, M.A. (eds.) Medical
  Imaging 2020: Computer-Aided Diagnosis. {SPIE} (Mar 2020)

\bibitem{Tang2019SPIE}
Tang, Y.B., Oh, S., Tang, Y.X., Xiao, J., Summers, R.M.: {CT}-realistic data
  augmentation using generative adversarial network for robust lymph node
  segmentation. In: Medical Imaging: Computer-Aided Diagnosis. vol. 10950, p.
  109503V (2019)

\bibitem{tang2018ct}
Tang, Y., Cai, J., Lu, L., Harrison, A.P., Yan, K., Xiao, J., Yang, L.,
  Summers, R.M.: {CT} image enhancement using stacked generative adversarial
  networks and transfer learning for lesion segmentation improvement. In:
  International Workshop on Machine Learning in Medical Imaging. pp. 46--54
  (2018)

\bibitem{tang2019xlsor}
Tang, Y., Tang, Y., Xiao, J., Summers, R.M.: {XLS}or: A robust and accurate
  lung segmentor on chest x-rays using criss-cross attention and customized
  radiorealistic abnormalities generation. In: International Conference on
  Medical Imaging with Deep Learning. pp. 457--467 (2019)

\bibitem{tang2019abnormal}
Tang, Y.X., Tang, Y.B., Han, M., Xiao, J., Summers, R.M.: Abnormal chest
  {X}-ray identification with generative adversarial one-class classifier. In:
  International Symposium on Biomedical Imaging. pp. 1358--1361 (2019)

\bibitem{tang2019tuna}
Tang, Y., Tang, Y., Sandfort, V., Xiao, J., Summers, R.M.: {TUNA-N}et:
  Task-oriented unsupervised adversarial network for disease recognition in
  cross-domain chest {X}-rays. In: International Conference on Medical Image
  Computing and Computer-Assisted Intervention. pp. 431--440 (2019)

\bibitem{tran2016edward}
Tran, D., Kucukelbir, A., Dieng, A.B., Rudolph, M., Liang, D., Blei, D.M.:
  {Edward: A library for probabilistic modeling, inference, and criticism}.
  arXiv e-prints: 1610.09787  (2016)

\bibitem{deeplesion}
Yan, K., Wang, X., Lu, L., Summers, R.M.: {DeepLesion: automated mining of
  large-scale lesion annotations and universal lesion detection with deep
  learning}. Journal of Medical Imaging  \textbf{5}(3),  1 -- 11 (2018)

\bibitem{taozhou2020}
{Zhou}, T., {Fu}, H., {Chen}, G., {Shen}, J., {Shen}, J., {Shao}, L.: Hi-net:
  Hybrid-fusion network for multi-modal {MR} image synthesis. IEEE Transactions
  on Medical Imaging pp.~1--1 (2020)

\bibitem{CycleGAN2017}
Zhu, J.Y., Park, T., Isola, P., Efros, A.A.: Unpaired image-to-image
  translation using cycle-consistent adversarial networks. In: International
  Conference on Computer Yision (ICCV). pp. 2223--2232 (2017)

\bibitem{zhu2020image}
Zhu, Y., Elton, D.C., Lee, S., Pickhardt, P., Summers, R.: Image translation by
  latent union of subspaces for cross-domain plaque detection. arXiv e-prints:
  2005.11384  (2020)

\bibitem{Zhu-2014-17155}
Zhu, Y., Huang, D., la~Torre~Frade, F.D., Lucey, S.: Complex non-rigid motion
  {3D} reconstruction by union of subspaces. In: Proceedings of CVPR (June
  2014)

\end{thebibliography}
%
% \begin{thebibliography}{8}
% \bibitem{ref_article1}
% Author, F.: Article title. Journal \textbf{2}(5), 99--110 (2016)
% 
% \end{thebibliography}
\end{document}